\documentclass[11pt,english]{article}
\usepackage{mathptmx}
\usepackage{courier}
\usepackage[T1]{fontenc}
\usepackage[latin9]{inputenc}
\usepackage{geometry}
\usepackage[active]{srcltx}
\usepackage{float}
\usepackage{amsmath}
\usepackage{graphicx}
\usepackage{esint}

\makeatletter

\newcommand{\lyxmathsym}[1]{\ifmmode\begingroup\def\b@ld{bold}
  \text{\ifx\math@version\b@ld\bfseries\fi#1}\endgroup\else#1\fi}

\makeatother

\usepackage{babel}
\begin{document}
\begin{center}
\textbf{\Large Orbital Angular Momentum in Scalar Diquark Model and
QED}\textbf{\date{}}
\par\end{center}

\begin{flushleft}
Hikmat BC%
\footnote{Presented by Hikmat BC at LIGHTCONE 2011, 23 - 27 May, 2011, Dallas.

}
\par\end{flushleft}

\begin{flushleft}
The University of Texas at El Paso, USA
\par\end{flushleft}

Email: hbbc@utep.edu,

\begin{flushleft}
Matthias Burkardt
\par\end{flushleft}

\begin{flushleft}
New Mexico State University, USA
\par\end{flushleft}

Email: burkardt@nmsu.edu

\renewcommand\abstractname{}
\begin{abstract}
We compare the orbital angular momentum of the \textquoteleft{}quark\textquoteright{}
in the scalar diquark model as well as that of the electron in QED
(to order $\alpha$) obtained from the Jaffe-Manohar decomposition
to that obtained from the Ji relation. We estimate the importance
of the vector potential in the definition of orbital angular momentum.\pagebreak{}
\end{abstract}

\section{{\large Introduction}}

~~~~~The question of how the total angular momentum $\frac{1}{2}$
of the proton is contributed by the spins of the quarks and gluons
and their orbital angular momenta, has been a puzzle since the European
Muon Collaboration (EMC) announced its result in late 80s. Very little
contribution to the proton's spin was found from the spin of the quarks,
and hence so- called proton {}``spin crisis'' has existed\cite{Myhrer:2009uq,Jaffe:1989jz}.
After almost 20 years of vigorous theoretical and experimental effort,
only about 30\% of the proton spin is contributed by spin of the quarks.
Researchers are actively engaged in to the quest for the remaining
70\% of the proton's spin. It appeared more clear that this rest of
the spin of the proton should be contributed by the orbital angular
momentum (OAM) of the quarks and gluons and the polarization of the
gluons. Recently, there are many debates on the proper way of decomposing
the total spin of the proton into the OAM and spin contribution from
quarks and gluons

\subsection{Ji Spin Sum Rule}

~~~~~Ji proposed a decomposition of the $\vec{z}$ - component
of the angular momentum of the nucleon

\begin{equation}
\frac{1}{2}=\frac{1}{2}\sum_{q}\Delta q+\sum_{q}L_{q}^{z}+J_{g}^{z}\label{eq:a-1}
\end{equation}

whose terms are matrix elements of the corresponding terms of the
$0xy$ - components of the following angular momentum tensor 

\begin{equation}
M^{0xy}=\frac{1}{2}\sum_{q}q^{\dagger}\Sigma^{z}q+\sum_{q}q^{\dagger}(\vec{r}\times i\vec{D})^{z}q+[\vec{r}\times(\vec{E}\times\vec{B})]^{z}
\end{equation}

where $i\vec{D}=i\vec{\partial}-g\vec{A}$. In this decomposition,
each term can be expressed as the expectation value of a manifestly
gauge invariant local operator. Also the total angular momentum of
the quark can be expressed in terms of the generalized parton distributions(GPDs)
as

\begin{eqnarray}
J_{q}^{z} & = & \frac{1}{2}\Delta q+L_{q}^{z}\nonumber \\
 & = & \frac{1}{2}\int_{0}^{1}dxx[q(x)+E_{q}(x,0,0)]
\end{eqnarray}

which can be measured in deeply virtual Compton scattering(DVCS) or
calculated in lattice gauge theory\cite{Burkardt:2008ua,Ji:1996ek,Ji:2002qa,Burkardt:2010wr}.

\begin{figure}
\begin{minipage}[t]{0.5\columnwidth}%
\includegraphics[scale=1.2]{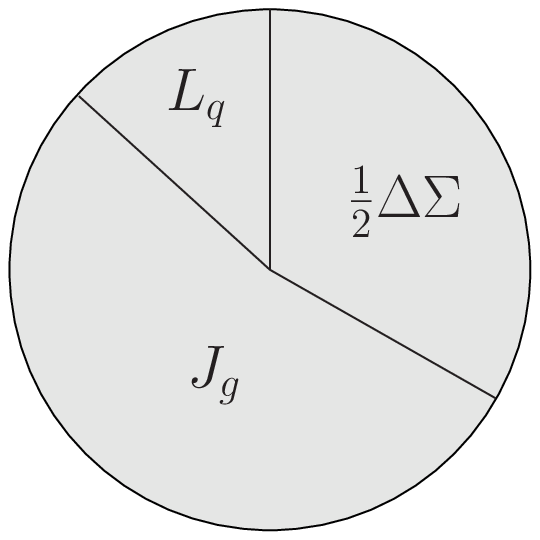}\caption{\label{fig:Ji-decomposition-1}Ji decomposition}
\end{minipage}%
\begin{minipage}[t]{0.5\columnwidth}%
\includegraphics[scale=1.2]{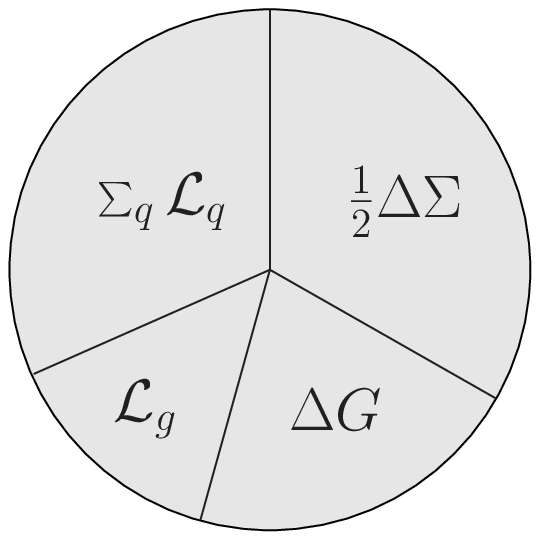}\caption{\label{fig:Jaffe--Manohar-decomposition-1}Jaffe- Manohar decomposition}
\end{minipage}
\end{figure}

\subsection{Jaffe - Manohar Spin Sum Rule}

~~~~~Jaffe and Manohar proposed a decomposition of the $\vec{z}$
- component of the angular momentum of the nucleon at the light -
cone frame as

\begin{equation}
\frac{1}{2}=\frac{1}{2}\sum_{q}\Delta q+\sum_{q}\mathcal{L}_{q}^{z}+\frac{1}{2}\Delta G+\mathcal{L}_{g}^{z}\label{eq:a}
\end{equation}

where these terms are defined as matrix elements of the corresponding
terms in the $+xy$ component of the angular momentum tensor

\begin{equation}
M^{+xy}=\frac{1}{2}\sum_{q}q_{+}^{\dagger}\gamma_{5}q_{+}+\sum_{q}q_{+}^{\dagger}(\vec{r}\times i\vec{\partial})^{z}q_{+}+\epsilon^{+-ij}TrF^{+i}A^{j}+2TrF^{+j}(\vec{r}\times i\vec{\partial})^{z}A^{j}\label{eq:b}
\end{equation}

where $q_{+}=\frac{1}{2}\gamma^{-}\gamma^{+}q$ is the dynamical component
of the quark field operators and $A^{+}\equiv A^{0}+A^{+}=0$ is the
light cone gauge.

In (\ref{eq:a}) and (\ref{eq:b}), the first and third terms are
the 'intrinsic' contributions to the angular momentum $J^{z}=$$+\frac{1}{2}$
of the nucleon and can be interpreted as spin of the quark and gluon
respectively and the second and third terms are interpreted as the
corresponding orbital angular momenta(OAM). The quark spin term is
manifestly gauge invariant. Gluon spin is accessible experimentally
and hence it is also gauge invariant. It is defined through a non
- local operator in the gauges other than light- cone gauge\cite{Burkardt:2008ua,Jaffe:1989jz,Burkardt:2010wr}.

The total OAM of both quark and gluon, which is gauge invariant, can
be written as

\begin{equation}
\mathcal{L}^{z}=\sum_{q}\mathcal{L}_{q}^{z}+\mathcal{L}_{g}^{z}=\frac{1}{2}-\frac{1}{2}\sum_{q}\Delta q-\frac{1}{2}\Delta G
\end{equation}

The expectation value of $\overline{q}\gamma^{z}\Sigma^{z}q$ vanishes
for a parity eigenstate. So, one can substitute $q^{\lyxmathsym{\dag}}\Sigma^{z}q\rightarrow\overline{q}\gamma^{+}\Sigma^{z}q=q_{+}^{\text{\dag}}\gamma_{5}q_{+}$,
i.e. the $\Delta q$ are same in both decompositions. All the other
terms are different from each other since they are not defined through
matrix elements of the same operator and one should not expect them
to have the same numerical value\cite{Burkardt:2008ua,Burkardt:2010wr}. 

We have OAM term from Ji relation,

\begin{equation}
q^{\dagger}(\vec{r}\times i\vec{D})^{z}q=\overline{q}\gamma^{0}(\vec{r}\times i\vec{D})^{z}q\rightarrow\overline{q}(\gamma^{0}+\gamma^{z})(\vec{r}\times i\vec{D})^{z}q+q_{+}^{\dagger}(\vec{r}\times i\vec{D})^{z}q_{+}\label{eq:c}
\end{equation}

Note that the expectation value is taken in a parity eigenstate. Even
in light- cone gauge, $\mathcal{L}^{z}$ and $L^{z}$ differ by the
expectation value of $q_{+}^{\dagger}(\vec{r}\times g\vec{A})q_{+}$
since Eq. (\ref{eq:c}) contains the transverse component of the vector
potential through the gauge covariant derivative.

\section{{\large Orbital Angular Momentum (OAM) in Scalar Diquark Model}}

~~~~~With the center of momentum and relative $\perp$ coordinates,
for a two particle system\cite{Burkardt:2008ua},

\begin{eqnarray}
\mathbf{P}_{\perp} & \equiv & \mathbf{p}_{1\perp}+\mathbf{p}_{2\perp}\\
\mathbf{R}_{\perp} & \equiv & x_{1}\mathbf{r}_{1\perp}+x_{2}\mathbf{r}_{2\perp}=x\mathbf{r}_{1\perp}+(1-x)\mathbf{r}_{2\perp}\nonumber \\
\mathbf{k}_{\perp} & \equiv & x_{2}\mathbf{p}_{1\perp}-x_{1}\mathbf{p}_{2\perp}=(1-x)\mathbf{p}_{1\perp}-x\mathbf{p}_{2\perp}\nonumber \\
\mathbf{r}_{\perp} & \equiv & \mathbf{r}_{1\perp}-\mathbf{r}_{2\perp}
\end{eqnarray}

where $x_{1}=x$ and $x_{2}=1-x$ are the momentum transfer carried
by active quark and the spectator respectively. One can replace the
OAM operator for particle 1 by $(1-x)$ times the relative OAM in
a state with $\mathbf{P}_{\perp}=0$ which gives us $\mathbf{p}_{1\perp}=-\mathbf{p}_{2\perp}=\mathbf{k}_{\perp}$
.

\begin{equation}
\mathcal{L}_{1}^{z}=\mathbf{r}_{1\perp}\times\mathbf{p}_{1\perp}=[\mathbf{R}_{\perp}+(1-x)\mathbf{r}_{\perp}]\times\mathbf{k}_{\perp}\rightarrow(1-x)\mathbf{r}_{\perp}\times\mathbf{k}_{\perp}=(1-x)\mathcal{L}^{z}
\end{equation}

Similarly, one can write $\mathcal{L}_{2}^{z}=x\mathcal{L}^{z}$ for
particle 2.

To compute the OAM of the quark in Ji and Jaffe and Manohar decomposition,
we have the light -cone Fock state wave functions in Scalar di quark
model\cite{Burkardt:2008ua,Brodsky:2000ii,Burkardt:2010wr},

\[
\Psi_{+\frac{1}{2}}^{\uparrow}(x,\overrightarrow{k_{\perp}})=(M+\frac{m}{x})\psi(x,\mathbf{k}_{\perp}^{2}),
\]

\begin{equation}
\Psi_{-\frac{1}{2}}^{\uparrow}(x,\overrightarrow{k_{\perp}})=-\frac{(k^{1}+ik^{2})}{x}\psi(x,\mathbf{k}_{\perp}^{2})\label{eq:35}
\end{equation}

\[
{|\psi|}^{2}=\frac{g^{2}x^{2}(1-x)}{[M^{2}x^{2}-(M^{2}+m^{2}-\lambda^{2})x+(\overrightarrow{k_{\perp}}^{2}+m^{2})]^{2}}
\]

where $g$ is the Yukawa coupling and $M$, $m$, and $\lambda$ are
the masses of the nucleon, quark and diquark respectively. Here $x$
is the momentum fraction carried by the quark and the relative momentum
$\mathbf{k}_{\perp}\equiv\mathbf{k}_{e\perp}-\mathbf{k}_{\gamma\perp}$.
The $\uparrow$ , upper index, of the wave function represents the
helicity of the nucleon and the lower index that of the quark.

According to Jaffe - Manohar decomposition, the OAM of the quark is\cite{Burkardt:2008ua}

\begin{equation}
{\cal L}_{q}^{z}=\frac{g^{2}}{16\pi^{3}}\int_{0}^{1}dx\int d^{2}\overrightarrow{k_{\perp}}(1-x){|\Psi_{\frac{-1}{2}}^{\uparrow}|}^{2}\label{eq:36}
\end{equation}

Similarly according to Ji decomposition,

\begin{equation}
L_{q}^{z}=J_{q}^{z}-\langle{S_{q}^{z}}\rangle
\end{equation}

\[
J_{q}^{z}=\frac{1}{2}\biggl(A_{q}(0)+B_{q}(0)\biggr)=\frac{1}{2}\int_{0}^{1}dxx[q(x)+E_{q}(x,0,0)]
\]

\[
A_{q}(0)=1-\frac{1}{16\pi^{3}}\iint{dx}{d^{2}k_{\perp}}(1-x)\biggl[{|\Psi_{+\frac{1}{2}}^{\uparrow}|}^{2}+{|\Psi_{-\frac{1}{2}}^{\uparrow}|}^{2}\biggr]
\]

\[
B_{q}(0)=\frac{1}{16\pi^{3}}\iint{dx}{d^{2}k_{\perp}}\frac{2M(1-x)(m+Mx)}{x}{|\psi|}^{2}
\]

\begin{equation}
\langle{S_{q}^{z}}\rangle=\frac{1}{2}+\frac{1}{16\pi^{3}}\iint{dx}{d^{2}k_{\perp}}\biggl(\frac{1}{2}\biggl[+1{|\Psi_{+\frac{1}{2}}^{\uparrow}|}^{2}-1{|\Psi_{-\frac{1}{2}}^{\uparrow}|}^{2}\biggr]-\frac{1}{2}\biggl[+1{|\Psi_{+\frac{1}{2}}^{\uparrow}|}^{2}+1{|\Psi_{-\frac{1}{2}}^{\uparrow}|}^{2}\biggr]\biggr)\label{eq:37}
\end{equation}

We used manifestly Lorentz invariant Pauli- Villars regularization
(subtraction with heavy scalar $\lambda^{2}\rightarrow\Lambda^{2}$)
to compute some of the divergent $\mathbf{k}_{\perp}$ integrals.
Computing above integrals we found that , in scalar diquark model,

\begin{equation}
L_{q}^{z}={\cal L}_{q}^{z}
\end{equation}

\begin{figure}[H]
\includegraphics[scale=1.2]{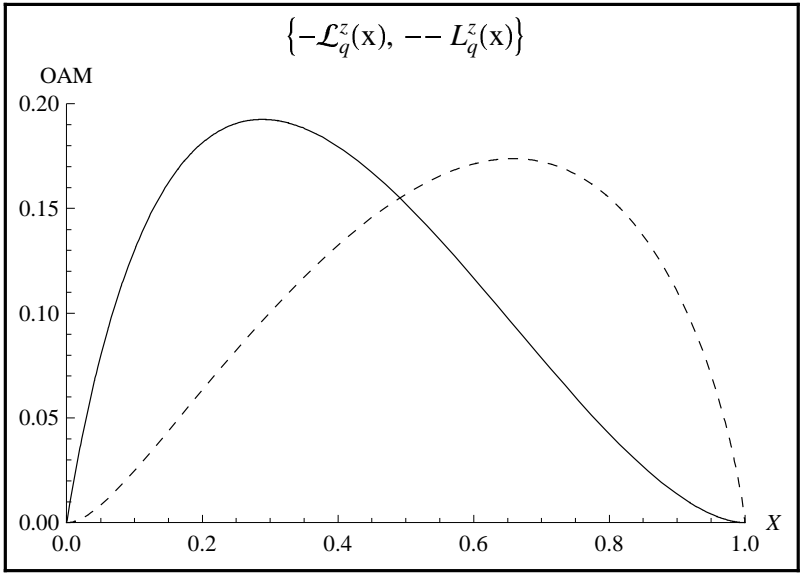}

\caption{\label{fig:x-dependence of OAM}$x$ dependence of OAM for Ji (dashed)
and Jaffe - Manohar (solid) in the scalar diquark model for parameters
$\Lambda^{2}=10\lambda^{2}=10m^{2}$. Both in units of $\frac{g^{2}}{16\pi^{2}}$
.}
\end{figure}

~~~~~It is not so surprising for scalar diquark model since it
is not a gauge theory i.e the OAM term does not contain a gauge field
term. However, the $x$ - distribution of the OAM, ($L_{q}^{z}(x)\text{ and }\mathcal{L}_{q}^{z}(x)$),
are not exactly some as shown in the Figure \ref{fig:x-dependence of OAM}.

\section{{\large Orbital Angular Momentum (OAM)  in QED}}

The light -cone Fock state wave functions in QED perturbative theory
are\cite{Brodsky:2000ii,Burkardt:2008ua,Burkardt:2010wr,Brodsky:1980zm}. There are
four polarization states in the $e\gamma$ Fock component :

\[
\Psi_{+\frac{1}{2}+1}^{\uparrow}(x,\overrightarrow{k_{\perp}})=-\sqrt{2}\frac{(-k^{1}+ik^{2})}{x(1-x)}\psi(x,\overrightarrow{k_{\perp}}^{2}),
\]

\[
\Psi_{+\frac{1}{2}-1}^{\uparrow}(x,\overrightarrow{k_{\perp}})=-\sqrt{2}\frac{(+k^{1}+ik^{2})}{(1-x)}\psi(x,\overrightarrow{k_{\perp}}^{2}),
\]

\[
\Psi_{-\frac{1}{2}+1}^{\uparrow}(x,\overrightarrow{k_{\perp}})=-\sqrt{2}(m-\frac{m}{x})\psi(x,\overrightarrow{k_{\perp}}^{2}),
\]

\begin{equation}
\Psi_{-\frac{1}{2}-1}^{\uparrow}(x,\overrightarrow{k_{\perp}})=0\label{eq:29}
\end{equation}

where
\[
\psi(x,\overrightarrow{k_{\perp}}^{2})=\frac{e/\sqrt{1-x}}{m^{2}-(\overrightarrow{k_{\perp}}^{2}+m^{2})/x-(\overrightarrow{k_{\perp}}^{2}+\lambda^{2})/(1-x)}
\]

where $x$ is the momentum fraction carried by the electron and $(1-x)$
is that for the photon. $\vec{k}_{\perp}$is the transverse component
of momentum of the electron; $m$ and $\lambda$ are the masses of
the electron and photon respectively.

According to Jaffe- Manohar decomposition, OAM of the electron,

\begin{equation}
{\cal \mathcal{L}}_{e}^{z}=\frac{1}{16\pi^{3}}\int_{0}^{1}{dx}\int{d^{2}k_{\perp}}(1-x)\biggl[-1{|\Psi_{+\frac{1}{2}+1}^{\uparrow}|}^{2}+1{|\Psi_{+\frac{1}{2}-1}^{\uparrow}|}^{2}\biggr]\label{eq:30}
\end{equation}

Similarly, according to Ji decomposition, OAM of the electron is 

\begin{equation}
L_{e}^{z}=\frac{1}{2}\biggl[A_{q}(0)+B_{q}(0)\biggr]-\langle S_{q}\rangle\label{eq:31}
\end{equation}

\begin{eqnarray}
A_{q}(0) & = & 1+\frac{1}{16\pi^{3}}\iint{dx}{d^{2}k_{\perp}}x\biggl[+1{|\Psi_{+\frac{1}{2}+1}^{\uparrow}|}^{2}+1{|\Psi_{+\frac{1}{2}-1}^{\uparrow}|}^{2}+1{|\Psi_{-\frac{1}{2}+1}^{\uparrow}|}^{2}\biggr]\nonumber \\
 &  & -\frac{1}{16\pi^{3}}\iint{dx}{d^{2}k_{\perp}}\biggl[+1{|\Psi_{+\frac{1}{2}+1}^{\uparrow}|}^{2}+1{|\Psi_{+\frac{1}{2}-1}^{\uparrow}|}^{2}+1{|\Psi_{-\frac{1}{2}+1}^{\uparrow}|}^{2}\biggr]\label{eq:32}
\end{eqnarray}

\begin{equation}
B_{q}(0)=4M\frac{1}{16\pi^{3}}\iint{dx}{d^{2}k_{\perp}}(m-mx){\psi|}^{2}\label{eq:33}
\end{equation}

Now, expectation value of spin angular momentum of the electron$\langle{S_{q}}\rangle$is 

\begin{eqnarray}
\langle{S_{q}}\rangle & = & \frac{1}{16\pi^{3}}\iint{dx}{d^{2}k_{\perp}}\frac{1}{2}\biggl[+1{|\Psi_{+\frac{1}{2}+1}^{\uparrow}|}^{2}+1{|\Psi_{+\frac{1}{2}-1}^{\uparrow}|}^{2}-1{|\Psi_{-\frac{1}{2}+1}^{\uparrow}|}^{2}\biggr]\nonumber \\
 &  & +\frac{1}{2}-\frac{1}{16\pi^{3}}\iint{dx}{d^{2}k_{\perp}}\frac{1}{2}\biggl[+1{|\Psi_{+\frac{1}{2}+1}^{\uparrow}|}^{2}+1{|\Psi_{+\frac{1}{2}-1}^{\uparrow}|}^{2}+1{|\Psi_{-\frac{1}{2}+1}^{\uparrow}|}^{2}\biggr]\label{eq:34}
\end{eqnarray}

We used manifestly Lorentz invariant Pauli- Villars regularization
(subtraction with heavy scalar $\lambda^{2}\rightarrow\Lambda^{2}$)
to compute some of the divergent $\mathbf{k}_{\perp}$ integrals.
Computing above integrals we found that\cite{Burkardt:2008ua,Burkardt:2010wr},

\begin{equation}
\mathcal{L}_{e}^{z}=-\frac{\alpha}{2\pi}\int_{0}^{1}dx(1-x^{2})\log\frac{(1-x)^{2}m^{2}+x\Lambda^{2}}{(1-x)^{2}m^{2}+x\lambda^{2}}\underrightarrow{\Lambda\rightarrow\infty,\lambda\rightarrow0}-\frac{\alpha}{4\pi}\left[\frac{4}{3}\log\frac{\Lambda^{2}}{m^{2}}-\frac{2}{9}\right]
\end{equation}

and 

\begin{eqnarray}
L_{e}^{z} & = & -\frac{\alpha}{4\pi}\int_{0}^{1}dx(1+x^{2})[\log\frac{(1-x)^{2}m^{2}+x\Lambda^{2}}{(1-x)^{2}m^{2}+x\lambda^{2}}-\frac{(1-x)^{2}m^{2}}{(1-x)^{2}m^{2}+x\lambda^{2}}-\nonumber \\
 &  & \frac{(1-x)^{2}m^{2}}{(1-x)^{2}m^{2}+x\Lambda^{2}}]
\end{eqnarray}

\begin{equation}
\underrightarrow{\Lambda\rightarrow\infty,\lambda\rightarrow0}\,\,\,\,\,-\frac{\alpha}{4\pi}\left[\frac{4}{3}\log\frac{\Lambda^{2}}{m^{2}}+\frac{7}{9}\right].
\end{equation}

As long as $\Lambda^{2}>\lambda^{2}$ , both the $\mathcal{L}_{e}^{z}$
and $L_{e}^{z}$ are negative regardless of the value of $\Lambda^{2}$.
So, the difference between two OAMs \cite{Burkardt:2008ua} is

\begin{equation}
\mathcal{L}_{e}^{z}-L_{e}^{z}\,\,\,\underrightarrow{(\Lambda\rightarrow\infty,\lambda\rightarrow0)}\,\,\,\frac{\alpha}{4\pi}
\end{equation}

\begin{figure}[H]
\includegraphics[scale=1.2]{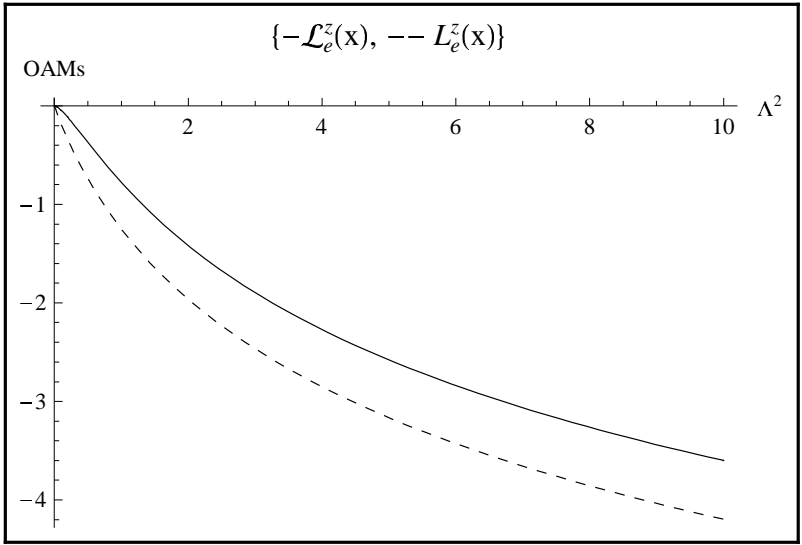}

\caption{\label{fig:Cutoff-dependence}Cutoff dependence of OAMs: Ji (dashed)
and Jaffe - Manohar (solid). Both in units of $\frac{\alpha}{4\pi}$
.}
\end{figure}

The cutoff dependence of OAMs from Ji and Jaffe - Manohar decompositions
in QED is shown in Figure \ref{fig:Cutoff-dependence}.

Above results allow us to evaluate the difference between two OAMs
for a massive quark \cite{Burkardt:2008ua} with $J^{z}=+\frac{1}{2}$,

\begin{equation}
\mathcal{L}_{q}^{z}-L_{q}^{z}=\frac{\alpha_{s}}{3\pi}
\end{equation}

In QCD, gluon OAM $\mathcal{L}_{g}^{z}$ is not experimentally accessible
but the gluon spin is. The total angular momentum of the gluon $J_{g}^{z}$
in the Ji relation is accessible either directly by calculating gluon
GPDs on a lattice and / or deeply virtual $J/\psi$ production or
indirectly by subtraction $J_{g}^{z}=\frac{1}{2}-J_{q}^{z}$ . One
can think of calculating OAM of gluon by subtracting $\frac{1}{2}\Delta G$
in Jaffe - Manohar decomposition from $J_{g}^{z}$ in Ji decomposition
but subtracting Eq. (\ref{eq:a}) from Eq. (\ref{eq:a-1}) gives us
\cite{Burkardt:2008ua},

\begin{equation}
J_{g}^{z}-\frac{1}{2}\Delta G=\mathcal{L}_{g}^{z}+\sum_{q}(\mathcal{L}_{q}^{z}-L_{q}^{z}).
\end{equation}

i.e numerically $J_{g}^{z}-\frac{1}{2}\Delta G$ differs from $\mathcal{L}_{g}^{z}$
by the same amount as $\sum_{q}\mathcal{L}_{q}^{z}$ differs from
$\sum_{q}L_{q}^{z}$.

In QED, the photon spin contribution in Ji decomposition is given
by

\begin{equation}
\Delta\gamma=\int_{0}^{1}dx\int\frac{d^{2}\vec{k_{\perp}}}{16\pi^{3}}\left[|\Psi_{+\frac{1}{2},+1}^{\uparrow}|^{2}-|\Psi_{+\frac{1}{2},-1}^{\uparrow}|^{2}+|\Psi_{-\frac{1}{2},+1}^{\uparrow}|^{2}\right]
\end{equation}

For $(\Lambda\rightarrow\infty,\lambda\rightarrow0)$ , we get,

\begin{equation}
J_{\gamma}^{z}-\frac{1}{2}\Delta\gamma=\mathcal{L}_{\gamma}^{z}+\frac{\alpha}{4\pi}.
\end{equation}

\section{{\large Summary and Discussion }}

~~~~~In this work, we studied the angular momentum decomposition
in scalar diquark model as well as that in QED as proposed by Jaffe
- Manohar and that according to Ji relation \cite{Burkardt:2010wr}.
Moreover, we compare OAMs of an electron in QED and that for an active
quark in scalar diquark model both in Jaffe- Manohar and Ji decompositions.
In the scalar diquark model, as anticipated, both the OAMs for the
fermions are same but not in QED. It can be concluded that the
presence of vector potential in the manifestly gauge invariant local
operator for the OAM does indeed contribute significantly to the numerical
value of the OAM. The differences seem to be small which are of the
order of $\frac{\alpha}{4\pi}$ but one should accept that the fact
that all the OAMs are of the order of $\alpha$. In QCD, for $\alpha_{s}\approx0.5$
about 10\% of the spin budget for the massive quark is also contributed
by vector potential term \cite{Burkardt:2008ua}. 

~~~~~~~~~~~~~~~~~~~~~~~~~~~~~~~~~~~~~~~~~~~~~~\_\_\_\_\_\_\_\_\_\_\_\_\_\_\_\_\_\_\_\_\_\_\_\_

\bibliographystyle{aip}
\bibliography{MYLibrary}

\end{document}